\documentclass[aps,prb,twocolumn,floatfix,superscriptaddress]{revtex4-1}
\usepackage{graphicx,txfonts,bm}
\usepackage[colorlinks,citecolor=magenta,linkcolor=blue]{hyperref}

\begin{document}


\title{Temperature dependence of correlated electronic states in archetypal kagome metal CoSn}


\author{Li Huang}
\email{lihuang.dmft@gmail.com}
\affiliation{Science and Technology on Surface Physics and Chemistry Laboratory, P.O. Box 9-35, Jiangyou 621908, China}

\author{Haiyan Lu}
\affiliation{Science and Technology on Surface Physics and Chemistry Laboratory, P.O. Box 9-35, Jiangyou 621908, China}

\date{\today}


\begin{abstract}
Hexagonal CoSn is a newly-discovered frustrated kagome metal. It shows close-to-textbook flat bands and orbital-selective Dirac fermions, which are largely associated with its strongly correlated Co-3$d$ orbitals. Because correlated electronic states are easily regulated by external conditions (such as chemical doping, pressure, and temperature), the fate of these kagome-derived electronic bands upon temperature becomes an interesting and unsolved question. In this work, we try to study the temperature-dependent electronic structures of hexagonal CoSn by means of the density functional theory in conjunction with the embedded dynamical mean-field theory. We find that hexagonal CoSn is in close proximity to a Mott insulating state at ambient condition. Special attention is devoted to the evolution of its Co-3$d$ electronic states with respect to temperature. At least six different temperatures (or energy scales), namely $T^{*}$, $T_{\text{FL}}$, $T_{\text{S1}}$ (and $T_{\text{S2}}$), $T_{\text{SF}}$, and $\bar{T}$, are figured out. They are related to stabilization of the ``pseudogap'' state, emergence of the non-Fermi-liquid phase, onset (and completeness) of the intermediate spin state, occurrence of the spin-frozen phase, beginning of the orbital freezing transition, respectively. 
\end{abstract}


\maketitle

\section{introduction\label{sec:intro}}

Recently, the $3d$-electron kagome metals have attracted lots of attentions~\cite{Ye2018,Yin2018,Liu2018,Ye2019,Wang_2020,Kang2020,PhysRevLett.121.096401,Yin2019,ws:2020,rc:2020,zl:2020}. In these quantum materials, 3$d$ transition metal (TM) atoms constitute layered kagome lattices, which are two-dimensional networks of corner-sharing triangles, resulting in exotic band topology~\cite{Ye2018}. On the one hand, this particular atomic arrangement gives rise to strongly localized TM-3$d$ electron wave-functions in real space. The corresponding electronic energy bands, which have extremely narrow bandwidths and are nearly dispersionless (so-called flat bands), naturally arise in momentum space through the destructive quantum interference mechanism~\cite{PhysRevB.45.12377,PhysRevB.34.5208,leykam}. That is one of the fingerprints of the kagome metals~\cite{Kang2020,rc:2020,Yin2019,PhysRevLett.121.096401,ws:2020}. On the other hand, crosses of symmetry-protected linearly dispersive bands (i.e., Dirac cones) are also one of the paradigmatic states of the kagome metals. Once spin-orbit coupling lying that lies in TM-3$d$ orbitals is nontrivial, considerable Dirac gaps will open and massive Dirac fermions will emerge~\cite{ws:2020,zl:2020,Yin2018}. Because of the unique combination of geometrically frustrated lattice symmetry and unusual band topology, the $3d$-electron kagome metals exhibit a great deal of distinguishing properties, including quantum spin-liquid states~\cite{PhysRevB.45.12377,RevModPhys.89.025003}, magnetic Weyl fermions~\cite{Kuroda2017,Wang2018}, and giant anomalous Hall effects~\cite{Liu2018,Wang2018,Kim2018}, just to name a few. Consequently, the 3$d$-electron kagome metals have been regarded as a versatile platform for studying the frustration-driven exotic spin-liquid phases, magnetic ground states, and novel topological excitations.

In these years, quite a few $3d$-electron kagome metals have been discovered. Their structural frameworks, in other words, the 2D kagome lattices, mostly comprise one of the following 3$d$ transition elements: Cr, Mn, Fe, Co, and Ni~\cite{Kang2020,Ye2018,PhysRevLett.121.096401,Ye2019,Wang_2020,Yin2018,Zhangeaao6791,Kim2018,Liu2018,Wang2018,Yin2019,PhysRevLett.124.077403,zl:2020,ws:2020,rc:2020}. Notice that the low-lying electronic states in these kagome metals, especially the flat bands and Dirac bands, are usually governed by the five TM-3$d$ orbitals~\cite{ws:2020}. Generally speaking, these TM-$3d$ orbitals are strongly correlated. There exist strong and orbital selective electronic correlations, which will lead to considerable band renormalization and orbital differentiation~\cite{georges:2013,PHILLIPS20061634,RevModPhys.70.1039}. The interplay of Coulomb repulsion interaction, Hund's exchange interaction, crystal-field splitting, and spin-orbit coupling makes the low-energy electronic states quite complicated, and finally contributes to the rich multi-orbital physics in the $3d$-electron kagome metals. 

There is no doubt that the electronic correlation should play a vital role in the electronic structures of 3$d$-electron kagome metals~\cite{ws:2020}. Firstly, the orbital energy levels are affected by the electronic correlation. As a consequence, the flat bands and Dirac bands are renormalized and shifted towards the Fermi level. Secondly, the TM-$3d$ electrons are redistributed among the five $d$-orbitals due to the electronic correlation, which lead to noticeable modifications of the orbital occupancies and the spin states. Finally, the strength of electronic correlation is easily tuned by external conditions, such as chemical doping, pressure, and temperature. The corresponding band structures are anticipated to be changed simultaneously~\cite{RevModPhys.70.1039}. For example, the correlated TM-3$d$ electrons should become more and more incoherent with increment of the system temperature. We wonder whether the representative flat bands and Dirac bands in the 3$d$-electron kagome metals could survive at moderately high temperature. In a word, we have to consider the electronic correlations in TM-3$d$ orbitals in order to gain a deep understanding about the electronic structures of 3$d$-electron kagome metals. But unfortunately, to our knowledge, the electronic correlation effect has not been fully taken into considerations in most of the previously theoretical and experimental researches concerning the 3$d$-electron kagome metals. Besides, the temperature effect is rarely counted.     

Now we would like to fill in this gap by investigating the temperature dependence of correlated electronic structures of an archetypal 3$d$-electron kagome metal CoSn. CoSn crystallizes in a hexagonal structure (space group $P_6/mmm$, No.~191), in which the kagome layers are stacked along the $c$-axis and separated by spacing layers [see Fig.~\ref{fig:akw}(a) and (b)]. The kagome layer is composed of a 2D kagome lattice of Co atoms and the centers of hexagons are occupied by Sn atoms (Co$_{3}$Sn), while the spacing layer is composed of a honeycomb lattice of Sn atoms (Sn$_{2}$) only. Very recently, the flat bands and Dirac bands in hexagonal CoSn have been determined both computationally and experimentally~\cite{ws:2020,rc:2020}. It is identified as an ideal 3$d$-electron kagome metal without complications induced by magnetism (the formation of magnetic ordering and local moments in this compound is suppressed presumably due to a higher $d$-orbital filling than the other 3$d$-electron kagome metals)~\cite{rc:2020} and with a perfect in-plane kagome lattice (the kagome lattice in hexagonal CoSn is the one closest to the 2D limit). Moreover, the magnitudes of energy gaps induced by spin-orbit coupling are quite different. They rely on the orbital characters of the Dirac bands, suggesting realization of orbital-selective Dirac fermions~\cite{ws:2020}. These findings shed new light on the multi-orbital physics in hexagonal CoSn and provide a reasonable explanation for the multiple topological electronic excitations in $3d$-electron kagome metals. 

In the present work, we tried to calculate the electronic structures of hexagonal CoSn at various temperatures by using a state-of-the-art first-principles many-body approach. The temperature-dependent momentum-resolved spectral functions, total and 3$d$ partial densities of states, self-energy functions, distributions of spin states, spin susceptibilities, and orbital susceptibilities were carefully evaluated. We find that though the Co-3$d$ states remain metallic, they are actually in the vicinity of a Mott-Hubbard transition. More importantly, the calculated results reveal a few characteristic temperatures (or equivalently, energy scales). They signal some furtive changes (including transitions and crossovers) in the electronic and spin states. An unified picture about how the correlated Co-3$d$ states in hexagonal CoSn evolve with the increment of temperature is finally provided.     

The rest of this paper is organized as follows. In section~\ref{sec:method}, we introduce the computational methods and important parameters. The main results are presented in section~\ref{sec:results}. Finally, section~\ref{sec:summary} serves as a short summary. 

\section{method\label{sec:method}}

A first-principles many-body approach, namely the density functional theory + embedded dynamical mean-field theory (dubbed DFT + eDMFT)~\cite{RevModPhys.78.865,RevModPhys.68.13,PhysRevB.81.195107}, was employed to calculate the electronic properties of hexagonal CoSn. This approach has been successfully applied to study the electronic structures of many strongly correlated materials, including transition metal oxides~\cite{PhysRevB.85.245110}, iron-based superconductors~\cite{Yin2011}, ruthenates~\cite{PhysRevB.86.195141,PhysRevB.100.125120,PhysRevLett.124.016401}, iridates~\cite{PhysRevLett.118.026404,PhysRevLett.111.246402}, and actinides~\cite{shim:2007,PhysRevB.99.045109}.        

For the DFT part, we used the \texttt{WIEN2K} code, which implements a full-potential linearized augmented plane wave formalism (FP-LAPW)~\cite{wien2k}. The exchange-correlation term in the Kohn-Sham equation was described within the generalized gradient approximation (actually the Perdew-Burke-Ernzerhof functional)~\cite{PhysRevLett.77.3865}. The muffin-tin radii for Co and Sn atoms are 2.46 and 2.42 a.u, respectively. We set $R_{\text{MT}} K_{\text{MAX}} = 8.0$ and used a $15 \times 15 \times 16$ $k$-mesh (245 $k$-points in the first irreducible Brillouin zone) for the Brillouin zone sampling.   

For the DMFT part, the Rutgers' \texttt{eDMFT} software package developed by K. Haule~\cite{PhysRevB.81.195107} was used. The correlated subspace includes the five Co-3$d$ orbitals. In order to define the DMFT projector which is used to project the Kohn-Sham bands into the local orbitals, a large energy window was used (from -10~eV to +10~eV with respect to the Fermi level). The Coulomb repulsive interaction parameter $U$ and Hund's exchange interaction parameter $J_{\text{H}}$ were 5.0~eV and 0.8~eV, respectively, which were borrowed from Reference~[\onlinecite{ws:2020}]. The rotationally-invariant-type Coulomb interaction was chosen in most DFT + eDMFT runs. A simplified Ising-type Coulomb interaction (only the density-density terms were included) was used in some benchmark tests. The obtained results are parallel and won't change the main conclusions of this paper. We employed the exact double counting scheme which is based on the dielectric constant approximation~\cite{PhysRevLett.115.196403} to cancel out the excess amount of the electronic correction effect that is included partly in the DFT part. We also compared our results with the ones that use the nominal double counting scheme~\cite{jpcm:1997}. The differences are negligible. In order to solve the auxiliary multi-orbital quantum impurity problems, the hybridization expansion version continuous-time quantum Monte Carlo impurity solver (dubbed CT-HYB)~\cite{RevModPhys.83.349,PhysRevB.75.155113} was employed. For the CT- HYB calculations, up to 200 million of Monte Carlo steps were employed for each quantum impurity solver run. In order to examine the temperature dependence of electronic structures of CoSn, we considered various $T$ from 60~K to 2400~K in the eDMFT calculations. We adopted the experimental lattice parameters and ignored the thermal expansion~\cite{ws:2020,rc:2020}. The system was assumed to be paramagnetic.    

We performed fully charge self-consistent DFT + eDMFT calculation~\cite{PhysRevB.81.195107}. About $60 \sim 80$ iterations were enough to obtain converged results. The convergent criteria for charge and energy were set to $1 \times 10^{-6}$e and $1 \times 10^{-6}$ Ry, respectively. Finally, the Matsubara self-energy functions were analytically continued from the imaginary to the real axis by the maximum entropy method~\cite{jarrell}. Then the real-frequency self-energy functions were utilized to calculate the other observables, such as quasiparticle band structures and density of states.  
 
\section{results and discussion\label{sec:results}}

\begin{figure*}[ht]
\centering
\includegraphics[width=\textwidth]{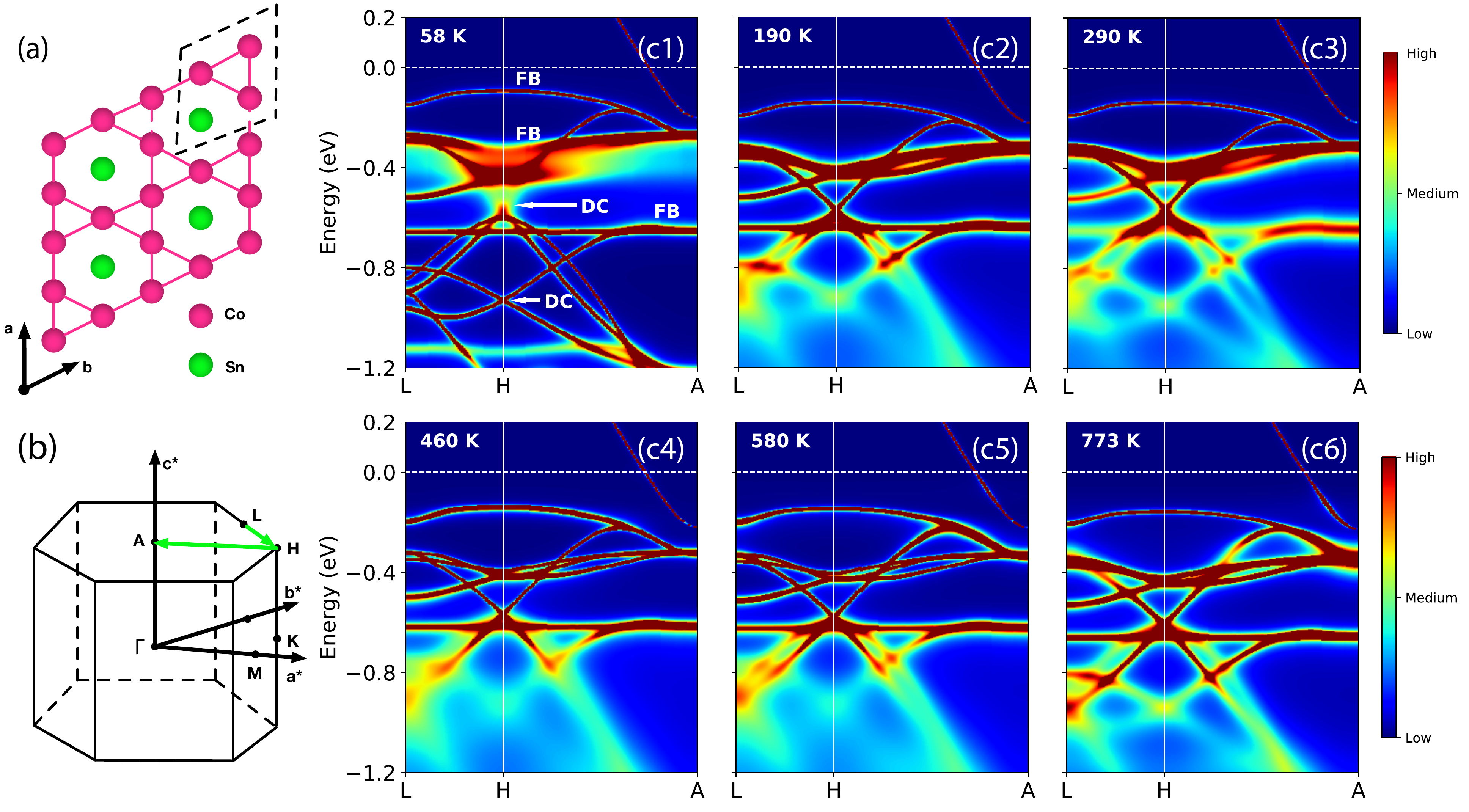}
\caption{(Color online). (a) Schematic picture for the kagome layer in hexagonal CoSn. The Co and Sn atoms are represented by red and green balls, respectively. The dashed rhomboid means the unit cell of the kagome lattice. (b) Irreducible Brillouin zone of hexagonal CoSn. Some high-symmetry points are labelled. Here the green arrows are used to depict the selected high-symmetry directions. (c1)-(c6) Momentum-resolved spectral functions $A(\mathbf{k},\omega)$ of hexagonal CoSn calculated by the DFT + DMFT method at different temperatures. The horizontal dashed lines denote the Fermi level. In panel (c1), the words ``FB'' and ``DC'' are abbreviations for flat bands and Dirac cones, respectively. \label{fig:akw}}
\end{figure*}

\begin{figure*}[ht]
\centering
\includegraphics[width=\textwidth]{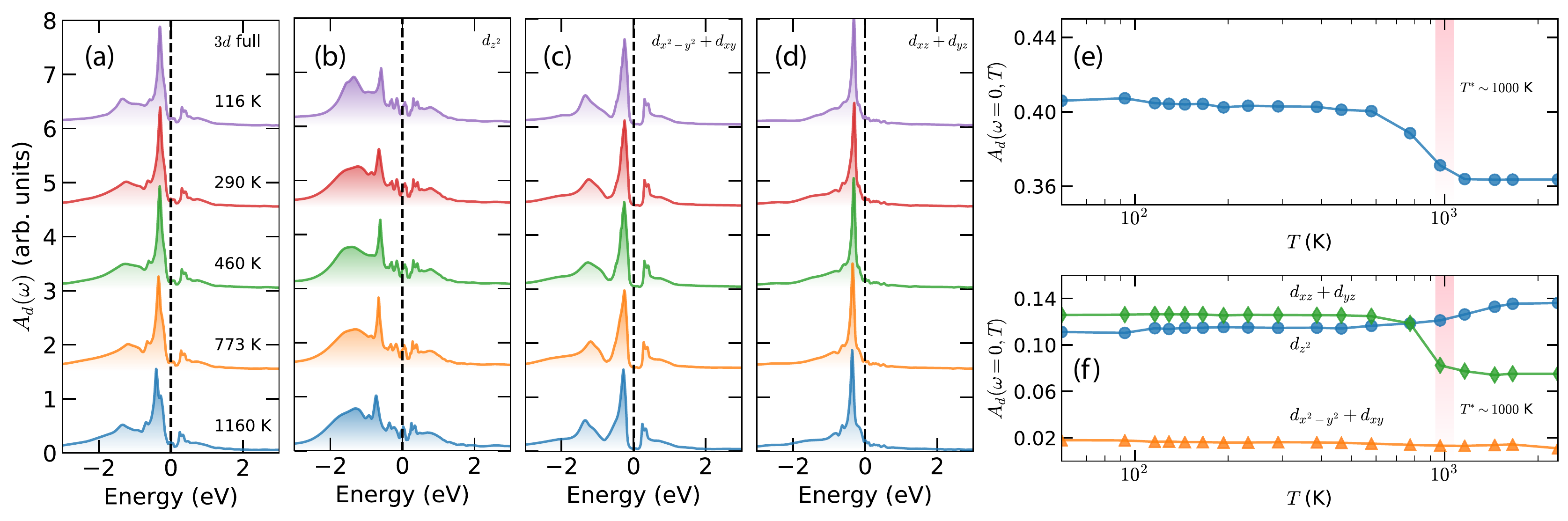}
\caption{(Color online). (a)-(d) Temperature-dependent electronic densities of states for various Co-3$d$ orbitals (full 3$d$, $d_{z^2}$, $d_{x^2-y^2} + d_{xy}$, and $d_{xz} + d_{yz}$ orbitals). The data shown in there panels have been rescaled for a better view. The vertical dashed lines denote the Fermi level. (e)-(f) Temperature dependences of spectral weights at $\omega = 0$ for various Co-3$d$ orbitals. In panels (e) and (f), the characteristic temperature $T^{*}$ ($\approx 1000$~K) is highlighted by using color bars. See text for explanations. \label{fig:dos}}
\end{figure*}

\begin{figure*}[ht]
\centering
\includegraphics[width=\textwidth]{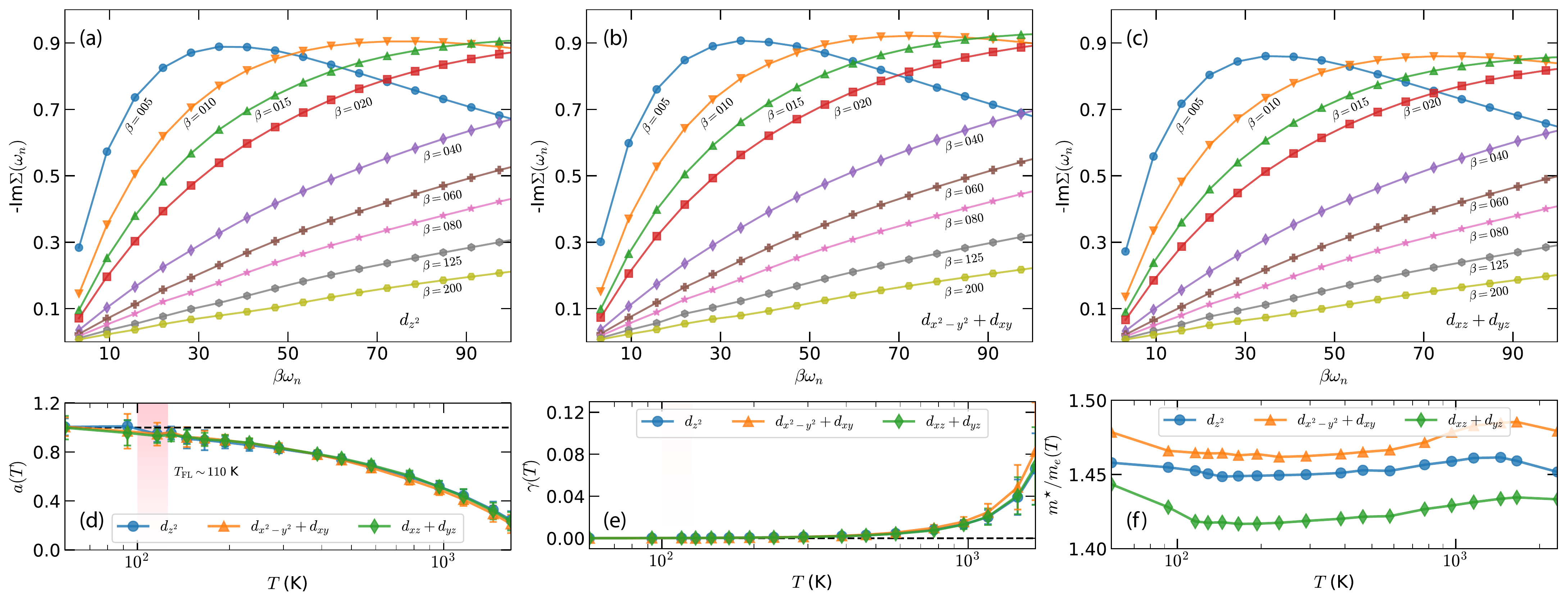}
\caption{(Color online). (a)-(c) Temperature-dependent Matsubara self-energy functions for various Co-$3d$ orbitals ($d_{z^2}$, $d_{x^2-y^2} + d_{xy}$, and $d_{xz} + d_{yz}$ orbitals). Here only the low-frequency imaginary parts are shown. $\beta$ means the inverse temperature ($\beta \equiv 1/T$). (d)-(e) Temperature dependences of orbital-resolved fitting parameters $a$ and $\gamma$. In panel (d), the horizontal dashed line denotes the ideal value of $a$ (i.e. $a \equiv 1.0$) predicted by the Landau Fermi-liquid theory. The characteristic temperature for Fermi-liquid state $T_{\text{FL}}$ is highlighted by using color bar. (f) Effective $3d$ electron masses estimated by using Eq.~(\ref{eq:mass}). See main text for more details. \label{fig:sig}}
\end{figure*}

\begin{figure*}[ht]
\centering
\includegraphics[width=\textwidth]{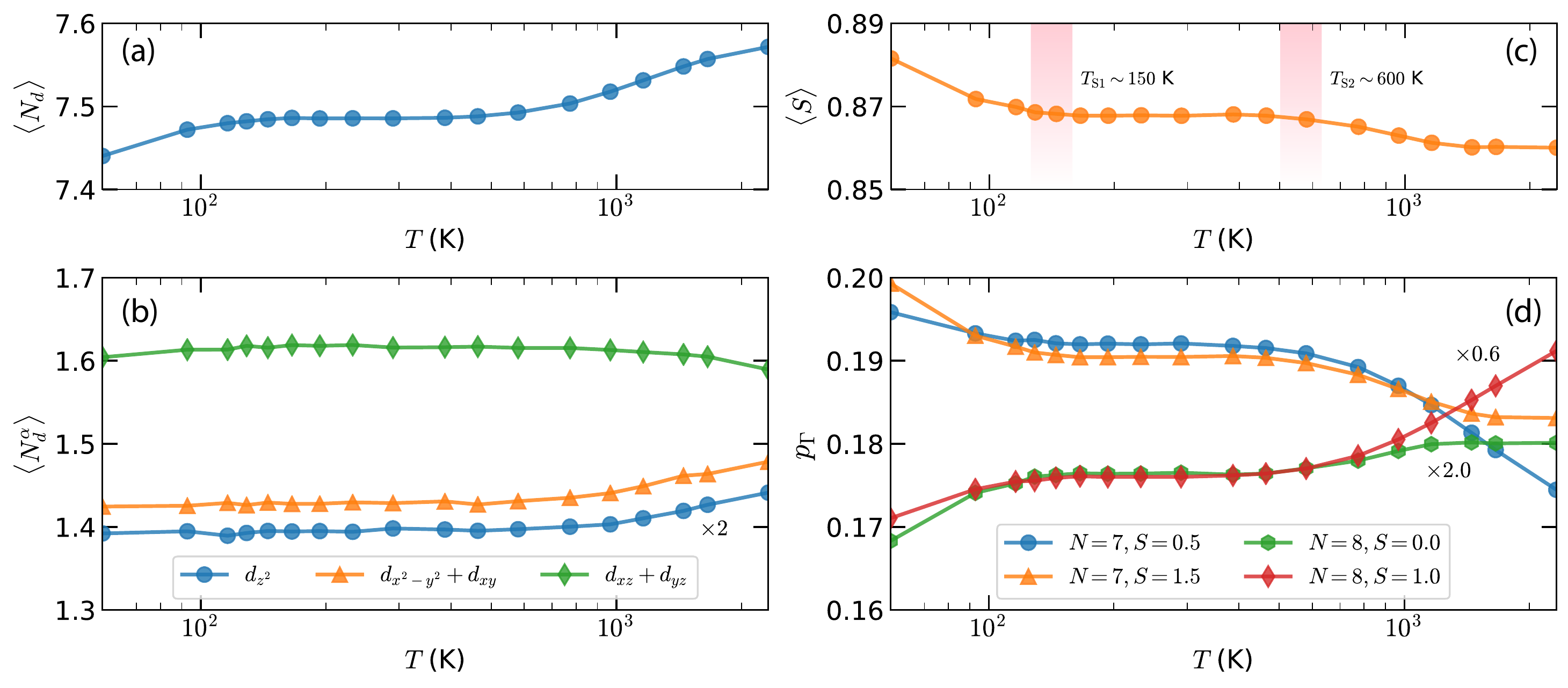}
\caption{(Color online). (a) Total 3$d$ occupancy and (b) orbital-resolved occupancies as a function of temperature $T$. (c) Total spin as a function of temperature $T$. $T_{\text{S1}}$ and $T_{\text{S2}}$ are two characteristic temperatures for possible spin-state crossovers. (d) Temperature dependences of probabilities for principal atomic eigenstates. Here, the data for atomic eigenstates $|N = 8, S = 0.0 \rangle$ and $|N = 8, S = 1.0 \rangle$ are rescaled for a better view. \label{fig:prob}}
\end{figure*}

\begin{figure*}[ht]
\centering
\includegraphics[width=\textwidth]{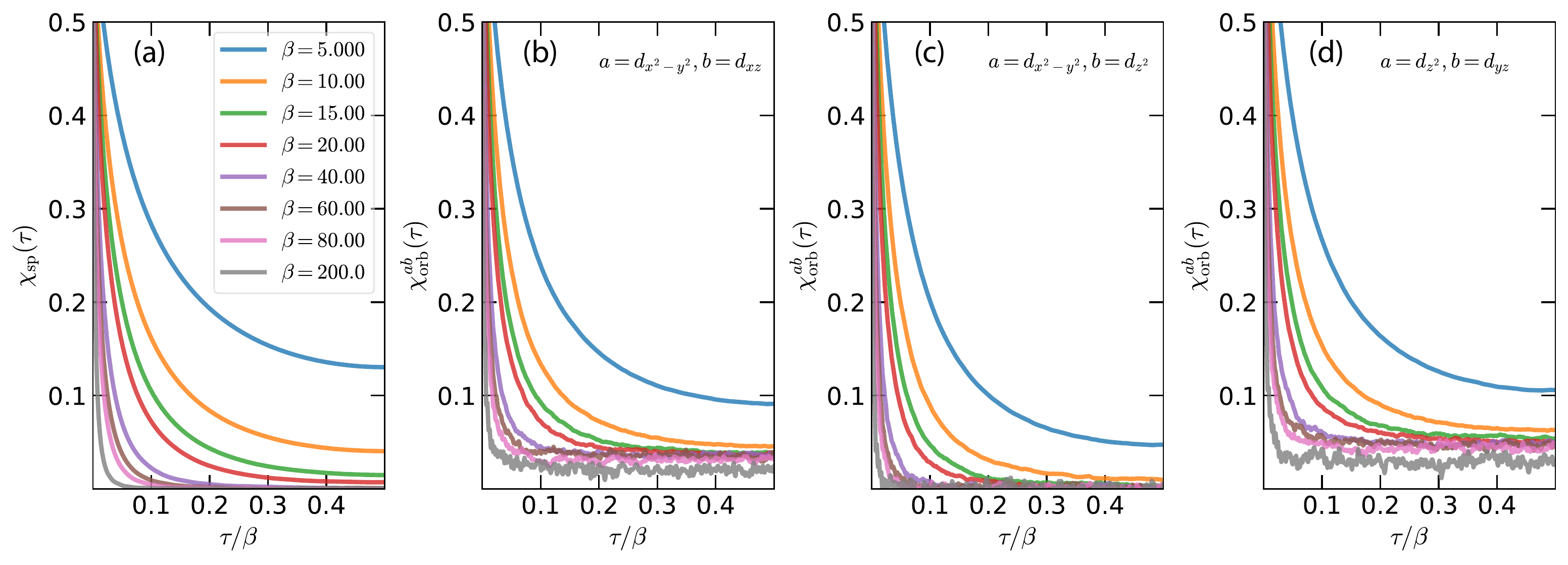}
\caption{(Color online). (a) Temperature-dependent spin-spin correlation functions $\chi_{\text{sp}}(\tau)$. (b)-(d) Temperature-dependent orbital-orbital correlation functions $\chi^{ab}_{\text{orb}}(\tau)$, which $a$ and $b$ are orbital indices. See main text for more details. \label{fig:susc}}
\end{figure*}

\subsection{Quasiparticle band structures}

We endeavored to calculate the temperature-dependent quasiparticle band structures (or equivalently, momentum-resolved spectral functions) $A(\mathbf{k},\omega)$ of hexagonal CoSn along some high-symmetry directions [$L-H-A$, please refer to Fig.~\ref{fig:akw}(b) for more details] at first. The calculated results for representative temperatures are visualized in Fig.~\ref{fig:akw}(c1)-(c6). Note that the quasiparticle band structures, Fermi surfaces, and surface states of hexagonal CoSn have been determined both experimentally (at $T = 60$~K and 20~K) and theoretically (at $T = 116$~K)~\cite{ws:2020,rc:2020}. Our calculated results are in excellent accord with them.  

All the essential features in the quasiparticle band structures are annotated using arrows and labels in Fig.~\ref{fig:akw}(c1). There are multiple flat bands (``FB'') between -0.6~eV and -0.1~eV. And there are two Dirac cones (``DC'') at $H$ point with binding energies $\sim$ 0.55 eV and $\sim$ 1.0~eV, respectively. These features suggest that the hexagonal CoSn is indeed a $3d$-electron kagome metal~\cite{ws:2020,rc:2020}. Even the system temperature is drastically increased, we find that the quasiparticle band structures are barely changed. The positions of the flat bands and Dirac bands are shifted slightly. But it might be due to the marginal effect of the DFT + eDMFT self-consistent iterations~\cite{PhysRevB.81.195107} or the biases introduced at the analytical continuation processes~\cite{jarrell}, instead of the temperature effect. Overall, the kagome-derived bands at low-energy region are quite stable and robust. They can survive at ultra-high temperature (at least up to 773~K). It seems that the correlated electronic states of hexagonal CoSn will not be changed greatly upon temperature. But it is not the case. Let us examine the other physical observables further. 

\subsection{Electronic density of states}

Figure~\ref{fig:dos}(a) shows the total $3d$ density of states $A_d(\omega)$. The following characteristics are revealed. (i) Since the 3$d$ spectral weights at the Fermi level are larger than zero [i.e. $A_d(\omega = 0) > 0.0$, see Fig.~\ref{fig:dos}(e) as well], the Co-$3d$ electronic states are strictly metallic. (ii) The quasiparticle resonance peaks are absent. Instead, there are ``pseudogap''-like structures in the Fermi level. Thus, it is concluded that the Co-3$d$ electronic states in CoSn are in the vicinity of Mott-Hubbard transition~\cite{RevModPhys.68.13,RevModPhys.70.1039}. In other words, CoSn could be easily tuned into a correlated insulator if the external conditions are changed. (iii) There are sharp peaks around -0.2~eV. They resemble the van-Hove singularities, which are associated with the flat bands as seen in Fig.~\ref{fig:akw}(c). We note that similar singularities have been observed in classic Hund's metals Sr$_{2}$RuO$_{4}$ and Sr$_{2}$MoO$_{4}$~\cite{PhysRevLett.124.016401,jk:2020}.   

Note that under hexagonal crystal field the five Co-3$d$ orbitals should be split into three groups, namely $d_{z^2}$, $d_{x^2-y^2} + d_{xy}$, and $d_{xz} + d_{yz}$. The two $d_{x^2-y^2} + d_{xy}$ orbitals are in-plane, while the three $d_{z^2}$ and $d_{xz} + d_{yz}$ orbitals are out-of-plane~\cite{ws:2020}. It is anticipated that their electronic structures could be quite different and present orbital differentiation physics. To validate this conjecture, we draw the orbital-resolved densities of states in Fig.~\ref{fig:dos}(b)-(d). For the $d_{z^2}$ orbital, multiple satellite peaks appear around the Fermi level. These peaks probably originate from the many-body transitions among various 3$d$ valence electron configurations (such as $3d^6$, $3d^7$, and $3d^8$), which are called quasiparticle multiplets sometimes~\cite{PhysRevB.101.195123,PhysRevB.81.035105}. This feature is usually seen in strongly correlated mixed-valence 4$f$ or 5$f$ electron systems~\cite{PhysRevB.101.195123,PhysRevB.81.035105}. It is quite surprising to discover it in correlated 3$d$-electron kagome metal. For the $d_{x^2-y^2} + d_{xy}$ orbitals, the spectrum shows a pseudogap between two side peaks. It looks as if the size (or width) of the pseudogap is not affected by the temperature effect. For the $d_{xz} + d_{yz}$ orbitals, the spectrum is characterized by a single sharp peak below the Fermi level (the side humps on the unoccupied side are somewhat small). Actually, the van-Hove singularities seen in Fig.~\ref{fig:dos}(a) are mainly from the contributions of the $d_{x^2-y^2} + d_{xy}$ and $d_{xz} + d_{yz}$ orbitals.

With increasing temperature, not only the shapes but also the peak positions of these spectra remain almost unchanged. That is consistent with what we have observed in the quasiparticle band structures [see Fig.~\ref{fig:akw}(c)]. However, considerable spectral weight transfer will take place at high temperature. Fig.~\ref{fig:dos}(e) shows the evolution of $A_d(\omega=0)$ with respect to temperature $T$. At low-temperature ($T < 600$~K) or high-temperature ($T >$ 1000~K) region, $A_{d}(\omega = 0,T)$ is approximately a constant. However, $A_{d}(\omega = 0,T)$ gradually decreases under temperature in the intermediate region, which signals that the Co-3$d$ electrons become more and more incoherent. Parts of Co-3$d$ valence electrons near the Fermi level would be excited into higher levels. As a consequence, the ``pseudogap'' state and the trend to a correlated insulator are greatly enhanced. So, we can define a new temperature scale $T^{*}$ ($\approx 1000$ K) to mark such a change in the spectral weight at zero frequency. Fig.~\ref{fig:dos}(f) displays the orbital-resolved $A_{d}(\omega = 0,T)$. The data for the $d_{x^2-y^2} + d_{xy}$ orbitals are featureless. Interestingly, the data for the $d_{z^2}$ and $d_{xz} + d_{yz}$ orbitals exhibit completely different behaviors. When $T < T^{*}$, the changes are rather small. When $T > T^{*}$, $A_d(\omega=0,T)$ of the $d_{z^2}$ orbital increases with increasing temperature, while those of the $d_{xz} + d_{yz}$ orbitals are on the contrary. Therefore, it is suggested that there is significant spectral weight transfer (or charge transfer) among the out-of-plane 3$d$ orbitals, while the in-plane $d_{x^2-y^2} + d_{xy}$ orbitals act as spectators only.           

\subsection{Self-energy functions}

It is well known that the electronic correlations in correlated electron systems are largely encapsulated in the self-energy functions~\cite{RevModPhys.68.13}. Thus, it is essential to inspect their properties. Fig.~\ref{fig:sig}(a)-(c) show the low-frequency parts of Matsubara self-energy functions of Co-$3d$ electrons (only the imaginary parts are presented here). When the temperature is low ($\beta > 40.0$, $T \approx 290$~K), they look like quasi-linear. However, when the temperature is high ($\beta \le 40.0$), they are convex functions. In order to describe this behavior more accurately, we applied the following equation to fit their low-frequency parts~\cite{PhysRevLett.101.166405}:
\begin{equation}
-\text{Im} \Sigma(\omega_n) = C (\omega_n)^{a} + \gamma. 
\end{equation}
The fitting parameters $a(T)$ and $\gamma(T)$ are shown in Fig.~\ref{fig:sig}(d)-(e). According to the Landau Fermi-liquid theory, $a = 1.0$ and $\gamma = 0.0$ denote the ideal Fermi-liquid state. Clearly, when the temperature is low, the system tends to obey the Fermi-liquid theory. On the contrary, when the temperature is high, the system shows an abnormal self-energy function that deviates from the description of Fermi-liquid theory, and enters the so-called non-Fermi-liquid region. Thus, we can define a new temperature scale again, $T_{\text{FL}}$, which signals the crossover from the Fermi-liquid state to the non-Fermi-liquid state. From Fig.~\ref{fig:sig}(d)-(e), we find the $T_{\text{FL}}$ is about 110~K for the hexagonal CoSn. Furthermore, it should be pointed out that the $T_{\text{FL}}$ has nothing to do with the orbital character. In other words, all Co-3$d$ orbitals share almost the same $T_{\text{FL}}$, regardless of whether they are in-plane or not.   

In correlated electron systems, the masses of interacting electrons should be renormalized. Therefore, the effective electron masses could be used as a valuable indicator to measure the strength of electron correlation. Next, we employed the following formula to estimate the effective masses of Co-3$d$ electrons $m^{\star}$~\cite{RevModPhys.68.13}:
\begin{equation}
\label{eq:mass}
Z^{-1} = \frac{m^{\star}}{m_e} \approx 1 - \frac{\text{Im} \Sigma(i\omega_0)}{\omega_0},
\end{equation}
where $\omega_0 (\equiv \pi / \beta)$ is the first fermionic Matsubara frequency, $Z$ is the quasiparticle weight, and $m_e$ is the mass of non-interaction electron. Notice that this formula is approximately correct at low temperature region~\cite{RevModPhys.68.13}. Fig.~\ref{fig:sig}(f) shows the calculated results. As a whole, $0.6 < Z < 0.8$, which indicates that the system is moderate correlated. We can see that the relationship $m^{\star}(d_{x^2-y^2} + d_{xy}) > m^{\star} (d_{z^2}) > m^{\star}(d_{xz} + d_{yz})$ holds for all temperatures. It means that the in-plane Co-$3d$ orbitals ($d_{x^2-y^2} + d_{xy}$ orbitals) are more correlated and suffer more renormalization than the out-of-plane orbitals ($d_{z^2}$ and $d_{xz} + d_{yz}$ orbitals). This explains why the ``pseudogap'' state occurs only at the $d_{x^2-y^2} + d_{xy}$ orbitals. The orbital differentiation and orbital selectivity are quite significant for the Co-3$d$ orbitals in hexagonal CoSn. It is possible to realize the orbital-selective Mott phase in this materials.

\subsection{Orbital occupancies and local spin states}

As mentioned above, $A_d(\omega = 0)$ will change with temperature, which might imply a temperature-driven redistribution of Co-3$d$ electrons. Here, we will provide some direct evidences about this issue. Fig.~\ref{fig:prob}(a) shows the total occupancy of Co-$3d$ orbitals as a function of temperature $T$. $\langle N_d \rangle$ increases with increasing temperature, which manifests that excess Co-$3d$ electrons may be from the weakly correlated Sn-$5p$ orbitals through the $p-d$ hybridization effect. The redistribution of electrons occurs not only between the Co-3$d$ orbitals and Sn-$5p$ orbitals, but also between the in-plane and out-of-plane Co-$3d$ orbitals. Fig.~\ref{fig:prob}(b) shows the temperature-dependent 3$d$ orbital occupancies $\langle N^{\alpha}_d\rangle$, where $\alpha$ denotes the orbital index. The $d_{xz} + d_{yz}$ orbitals will lose a small portion of electrons at high temperature. On the contrary, the $d_{x^2-y^2} + d_{xy}$ and $d_{z^2}$ will gain more electrons. This trend is roughly consistent with what we have learnt from the temperature dependence of $A_d(\omega = 0)$ [see Fig.~\ref{fig:dos}(e) and (f)]. Furthermore, we notice that over a wide range of temperature, the orbital occupancies look like being fixed. Both $\langle N_d \rangle$ and $\langle N^{\alpha}_d \rangle$ exhibit a platform in this temperature region. The widths of these platforms are the same, irrespective of the orbital characters. 

Since the spin states of the systems are in tightly connected with the orbital occupancies, it is supposed that the spin states of Co-3$d$ electrons will be modified as well. Fig.~\ref{fig:prob}(c) shows the expected values of total spin $\langle S \rangle$. Initially, it is in the ``high'' spin state, $\langle S \rangle \approx 0.88$. Then it decreases quickly with temperature until $T$ reaches $T_{\text{S1}}$. When $T_{\text{S1}} < T < T_{\text{S2}}$, it is in the intermediate spin state ($\langle S \rangle \approx 0.87$), and exhibits weak temperature dependence. Once $T$ is larger than $T_{\text{S2}}$, $\langle S \rangle$ decreases monotonously with temperature again, and the system enters the ``low'' spin state. Hence, we can define two new temperature scales, namely $T_{\text{S1}}$ and $T_{\text{S2}}$. They are used to locate the onset and finish of the intermediate spin state. In this material, $T_{\text{S1}}$ and $T_{\text{S2}}$ are about 150~K and 600~K, respectively. The pressure-driven high-spin to low-spin transition has been suggested for the cobalt monoxide~\cite{PhysRevB.85.245110}. In the spinel compound Co$_{3}$O$_{4}$, both the high-spin Co$^{2+}$ and the low-spin Co$^{3+}$ ions can coexist~\cite{PhysRevB.99.104104}. The underlying mechanisms for these phenomena have been well understood. Now we would like to clarify the underlying mechanism for this temperature-driven spin state transition in hexagonal CoSn. At first, we used some good quantum numbers, such as the total occupancy $N$ and the total spin $S$ to classify the atomic eigenstates $|\Gamma\rangle$ of the local impurity Hamiltonian $H_{\text{loc}}$ for the Co-3$d$ electrons. And then we tried to measure the atomic eigenstate probabilities $p_{\Gamma}$ via the CT-HYB quantum impurity solver~\cite{PhysRevB.75.155113,shim:2007}. Fig.~\ref{fig:prob}(d) illustrates the calculated results for some principal atomic eigenstates. Apparently, in the low temperature (``high'' spin state) region, the atomic eigenstates $|N = 7, S = 0.5 \rangle$ and $|N = 7, S = 1.5 \rangle$ dominate. However, in the high temperature (``low'' spin state) region, the atomic eigenstate $| N = 8, S = 1.0 \rangle$ is more favorable. So, it is the $|N = 7, S = 0.5 \rangle$ + $|N = 7, S = 1.5 \rangle$ $\to$ $| N = 8, S = 1.0 \rangle$ transition that results in the evolution of spin states. 

\subsection{Spin and orbital dynamics}

Next, let us focus on the spin dynamics of hexagonal CoSn. We tried to calculate the spin susceptibility $\chi_{\text{sp}}(\tau)$ for Co-3$d$ orbitals via the following definition:
\begin{equation}
\chi_{\text{sp}}(\tau) = \langle S(0) S(\tau) \rangle. 
\end{equation}
Here, $S$ is the operator of total spin, and $\tau$ denotes the imaginary time ($\tau \in [0,\beta]$). Fig.~\ref{fig:susc}(a) shows the calculated results. When the temperature is low, $\chi_{\text{sp}}(\tau)$ approaches zero for times $\tau$ sufficiently far from $\tau = 0$ or $\beta$. On the contrary, when the temperature is high, the asymptotic behavior of $\chi_{\text{sp}}(\tau)$ is quite different. It approaches a nonzero constant $c$ at large enough $\tau$, which means a well-defined frozen local moment. On the other hand, if the Landau Fermi-liquid theory is obeyed, $\chi_{\text{sp}}(\tau)$ should behave as $\chi_{\text{sp}}(\tau) \sim [T / \text{sin}(T\tau\pi)]^2$. Clearly, the asymptotic behavior of $\chi_{\text{sp}}(\tau)$ provides another evidence for the violation of the Landau Fermi-liquid theory at high temperature~\cite{PhysRevLett.101.166405}. Thus, we can define a new temperature scale $T_{\text{SF}}$. When $T < T_{\text{SF}}$, $\chi_{\text{sp}}(\tau = \beta/2) \to 0.0$. When $T > T_{\text{SF}}$, $\chi_{\text{sp}}(\tau = \beta/2) \to c$. It seems that the spin moment is frozen at $T > T_{\text{SF}}$. This situation is the so-called spin-freezing state. Indeed, $T_{\text{SF}}$ signals the emergence of a spin-freezing phase. For the hexagonal CoSn, its $T_{\text{SF}}$ is around 290~K ($\beta = 40$) according to Fig.~\ref{fig:susc}(a).

Then, another question is raised. How about the orbital dynamics? In order to answer this question, we computed the orbital susceptibility as well. The definition of orbital susceptibility $\chi^{ab}_{\text{orb}}(\tau)$ is as follows~\cite{deng:2019,STADLER2019365}:
\begin{equation}
\chi^{ab}_{\text{orb}}(\tau) = \langle O^{ab}(0) O^{ab}(\tau) \rangle,
\end{equation}
where
\begin{equation}
O^{ab} = n_{a} - n_{b}.
\end{equation}
Here $n_{a}$ (or $n_{b}$) denotes the occupancy of orbital $a$ (or $b$). In the present work, we only considered three typical combinations of $a$ and $b$: (i) $(a,b) = (d_{x^2 - y^2},d_{xz})$. (ii) $(a,b) = (d_{x^2 - y^2},d_{z^2})$. (iii) $(a,b) = (d_{z^2},d_{yz})$. The calculated results are depicted in Fig.~\ref{fig:susc}(b)-(d). For cases (i) and (iii), $\chi^{ab}_{\text{orb}}(\tau = \beta/2)$ is always larger than zero, regardless of the inverse temperature $\beta$. Similar to the spin-freezing phase, we call this behavior orbital-freezing state. It should be related with some kind of orbital orders. For case (ii), when $T > \bar{T}_{ab}$, $\chi^{ab}_{\text{orb}}(\tau = \beta/2)$ will show analogous behaviors to those as seen in cases (i) and (iii). But when $T < \bar{T}_{ab}$, it will approach zero. It is suggested that the orbital-freezing state will be destroyed below $\bar{T}_{ab}$. In this case, $\bar{T}_{ab} \gg T_{\text{SF}}$. Obviously, $\bar{T}_{ab}$ marks the temperature scale for the orbital-freezing transition. According to the calculated results, orbital freezing is not a universal and indiscriminate feature for all orbitals. It only occurs for special combinations of orbitals at high enough temperature.    

\section{Concluding remarks\label{sec:summary}}

In this paper, we present a systematic study about the temperature dependence of electronic structures of frustrated kagome materials CoSn. Though the hexagonal CoSn is an archetypal kagome metal, we find it is rather close to a correlated Mott insulator. Both the ``pseudogap'' and van-Hove singularity are observed in its band structures and densities of states, respectively. In addition, a series electronic and spin state transitions or crossovers are predicted. We figured out at least six different temperatures or energy scales, namely $T^{*}$, $T_{\text{FL}}$, $T_{\text{S1}}$ and $T_{\text{S2}}$, $T_{\text{SF}}$, and $\bar{T}_{ab}$, which are in connection with the ``pseudogap'', non-Fermi-liquid state, intermediate spin state, spin-freezing state, and orbital-freezing state of Co-3$d$ electrons, respectively. We established that $T^{*} \approx \bar{T}_{ab} \gg T_{\text{S2}} \gg T_{\text{SF}} > T_{\text{S1}} > T_{\text{FL}}$. It means that with the increment of temperature, the Co-3$d$ electrons in hexagonal CoSn should undergo the following changes (transitions or crossovers) successively: from the Fermi-liquid state to the non-Fermi-liquid state, entering the intermediate spin state, spin-freezing transition, entering the ``low'' spin state, orbital-freezing transition, and entering the ``pseudogap'' state. The calculated results suggest that the correlated Co-$3d$ electronic states in the hexagonal CoSn will be dramatically tuned by temperature. The temperature dependence of electronic structures of hexagonal CoSn is much more complex than what we have expected before and should be taken into consideration seriously. 

Finally, we speculate that similar properties could be detected in the other 3$d$-electron kagome metals. We would like to note that most of the ``hidden'' changes of the correlated electronic states presented in this paper occur in a two-particle level and at high temperature region. It is not an easy task to validate them experimentally. Anyway, our results shed new light into the electronic structures of strongly correlated 3$d$-electron kagome metals. It would be essential and interesting to examine the temperature-dependent electronic structures of the other strongly correlated metals, such as iron-based superconductors, ruthenates, iridates, and actinides.

\begin{acknowledgments}
This work was supported by the Natural Science Foundation of China (No.~11874329, 11934020, and 11704347), and the Science Challenge Project of China (No.~TZ2016004).
\end{acknowledgments}


\bibliography{cs}

\end{document}